\documentclass[12pt,a4paper]{article}
\usepackage{jheppub}

\title{
Twisted bosonization in two dimensional noncommutative spacetime}

\author{Asrarul Haque}
\author{and T.R. Govindarajan}

\affiliation{Institute of Mathematical Sciences,\\
 CIT Campus, Taramani, Chennai 600113
(INDIA)}
\emailAdd{asrarul@imsc.res.in,trg@imsc.res.in}

\abstract{We study the twisted bosonization of massive Thirring model to relate to
sine-Gordon model in Moyal spacetime using twisted commutation relations. We obtain the 
relevant twisted bosonization rules. 
We show that there exists dual relationship between twisted bosonic and fermionic operators.
The strong-weak duality is also observed to be preserved as its commutative counterpart.}


\begin{document}
\maketitle  

\section{Introduction}
Quantum field theories in 1+1 dimensional commutative spacetime exhibit novel phenomenon of bosonization. Bosonization maps the bosonic composite operators to the
 fermionic ones and vice versa.
Bosonization is expected in two dimensional quantum field theories as the phase change under exchange of fermions can be absorbed in the bosonic field operators
. The equivalence between the fermionic and the bosonic systems in two dimensions was discovered long ago
by Jordan and Wigner \cite{wig}. This was later demonstrated in continuum quantum field theories on equivalence between massive Thirring model 
and sine-Gordon model \cite{Sid,man}.
 The fermion-boson equivalence could be established as follows: The canonical quantization of a bosonic field operator in 1+1 dimensions:
$$\phi(x) = \int{\frac{dp^1}{\sqrt{2\pi}2p^0}[a(p)e^{-ip.x}+a^{\dagger}(p)e^{ip.x}]};~~~~~p^{\mu} = (p^0, p^1)$$
demands that the bosonic operators $a(p)$ and $a^{\dagger}(p)$ must satisfy the canonical commutation relations:
$$[a(p), a^{\dagger}(p') ] = 2p^0\delta(p-p');~~[a(p), a(p') ] = 0 = [a^{\dagger}(p), a^{\dagger}(p') ]$$
One can construct an operators $b(p)$ and $b^{\dagger}(p)$ in terms of $a(p)$ and $a^{\dagger}(p)$ defined by,
\begin{eqnarray}
 b^{\dagger}(p) &=& a^{\dagger}(p) e^{-i\pi\int_{p^1}^\infty\frac{dk^1}{2k^0}a^{\dagger}(k)a(k) }\\
 b(p) &=& e^{i\pi\int_{p^1}^\infty\frac{dk^1}{2k^0}a^{\dagger}(k)a(k) }a(p)
\end{eqnarray}
which are fermionic operators that satisfy the canonical anti-commutation relations:
$$\{b(p), b^{\dagger}(p')\} = 2p^0\delta(p-p');~~\{b(p), b(p') \} = \{b^{\dagger}(p), b^{\dagger}(p') \} = 0$$
The physical import of bosonization in a quantum field theory is understood in terms of the duality between the strong and weak coupling limits.
Consider the sine-Gordon model defined by the action:
\begin{equation}S_{SG}=\int{d^2x} (\frac{1}{2} \partial_{\mu}\phi\partial^{\mu}\phi + \frac{\alpha_0}{\beta^2}\cos\beta\phi+\gamma_0)\end{equation}
and massive Thirring model by,
\begin{equation} S_{MT} =\int{d^2x}(\bar \psi(i\not\partial -m_0)\psi -\frac{1}{2}g j^{\mu}j_{\mu})\end{equation}
These are dual to each other when coupling constants are related by $\frac{4\pi}{\beta^2} = 1+ \frac{g}{\pi}$ by the bosonization rules. 
Thus strong-weak duality implies that a weak bosonic coupling could be found for strong fermionic interactions.
In this paper we wish to address the following question: Can the two dimensional quantum field theories, in particular,
quantum massive Thirring model in Moyal spacetime be bosonized and obtain sine-Gordon model in the same spacetime? \vspace{0.1cm}

Moyal spacetime in 1+1 dimensions is defined by 
\begin{equation}
 [\hat x_{\mu},\hat x_{\nu}] = i\theta_{\mu\nu}\equiv i\epsilon_{\mu\nu}\theta ~~~~\mu,\nu = 0, 1 \label{nc1}.
\end{equation}
In the case of two dimensional spacetime, every antisymmetric matrix equals a
number times a constant antisymmetric second-rank tensor $\epsilon ^{\mu\nu}$ with  $\epsilon ^{12}=1$, which is Lorentz invariant. Therefore, we can write
$\theta^{\mu\nu} = i\theta \epsilon^{\mu\nu}$, where $\theta$ is a real parameter. The commutator (\ref{nc1}) is invariant
under translation:~$ x^\mu \to
x'^\mu=x^\mu+a^\mu,~ a^\mu \in \mathbb{R}$.
While Poincare group
 can be implemented in 1+1 dimensions due to covariance of $\epsilon_{\mu\nu}$, implementation of twisted Poincare group has played the role of an interesting 
development in the study of noncommutative quantum field theories \cite{mas0,bal1,bal4}.
Bosonization on noncommutative spacetime
in the context of conventional canonical quantization
have been studied in the past in 1+1 dimensions \cite{bos0,bos1} as well as in 2+1 dimensions \cite{bos2}. Integrable sine-Gordon model in Moyal 
spacetime using conventional quantization
has been studied in \cite{tr0,tr1,tr2}. Moreover, the ref. \cite{tr1}, also discusses bosonized NC massive Thirring model. However noncommutative quantum
field theories under conventional quantization, unlike twisted quantized QFT's, suffer from IR-UV mixing and unitarity difficulties.
Hence our program in this paper assumes importance. Interestingly the quantum
sinh-Gordon model is shown to be integrable using twisted 
quantization program on noncommutative space \cite{sac0}.
\vspace{0.1cm}

Twisted quantization program is required for implementing twisted Poincare symmetry in the quantum field theory on noncommutative
spacetime \cite{mas,sac1,sac2}. By twisted Poincare group one implies Poincare group implemented on multiparticle states through Drinfeld twist \cite{mas}. 
Now the following question naturally emerges: does there exist a twisted bosonic representation of the twisted fermionic operator
in the framework of quantized field operators in terms of the twisted creation and annihilation operators? Twisted
quantization program on Moyal spacetime is obtained through a deformation of the canonical quantization. To this end we consider the twisted bosonic operators
$a^{\theta}(p)$ and $a^{\theta\dagger}(p)$ that satisfy the commutation relations:
\begin{eqnarray*}
&& a^{\theta}(p) a^{\theta\dagger}(p') = e^{-ip\wedge p'}a^{\theta\dagger}(p')a^{\theta}(p) +  2p^0\delta(p-p');\\
&&a^{\theta}(p) a^{\theta}(p') = e^{ip\wedge p'}a^{\theta}(p')a^{\theta}(p);~~a^{\theta\dagger}(p) a^{\theta\dagger}(p')
 = e^{ip\wedge p'}a^{\theta\dagger}(p')a^{\theta\dagger}(p)\end{eqnarray*}
 where $p\wedge p' = p_{\mu}\theta^{\mu\nu}p'_{\nu} = p_{\mu}p'_{\nu}\epsilon^{\mu\nu}\theta$. 
One can construct the twisted operators $b^{\theta}(p)$ and $b^{\theta\dagger}(p)$ defined by,
\begin{eqnarray}
 b^{\theta\dagger}(p) &=& a^{\theta\dagger}(p) e^{-i\pi\int_{p^1}^\infty\frac{dk^1}{2k^0}a^{\theta\dagger}(k)a^{\theta}(k) } = a^{\theta\dagger}(p) e^{-i\pi\int_{p^1}^\infty\frac{dk^1}{2k^0}a^{\dagger}(k)a(k) }\\
 b^{\theta}(p) &=& e^{i\pi\int_{p^1}^\infty\frac{dk^1}{2k^0}a^{\theta\dagger}(k)a^{\theta}(k) }a^{\theta}(p) = e^{i\pi\int_{p^1}^\infty\frac{dk^1}{2k^0}a^{\dagger}(k)a(k) }a^{\theta}(p)
\end{eqnarray}
which are twisted fermionic operators that obey the anticommutation relations (see Appendix A):
\begin{eqnarray*}
&& b^{\theta}(p) b^{\theta\dagger}(p') =- e^{-ip\wedge p'}b^{\theta\dagger}(p')b^{\theta}(p) +  2p^0\delta(p-p');\\
&&b^{\theta}(p) b^{\theta}(p') = -e^{ip\wedge p'}b^{\theta}(p')b^{\theta}(p);~~b^{\theta\dagger}(p) b^{\theta\dagger}(p')
 = -e^{ip\wedge p'}b^{\theta\dagger}(p')b^{\theta\dagger}(p)\end{eqnarray*}
Thus the bosonic representation for the fermionic operator holds true even in the case of twisted quantization.\vspace{0.1cm}

In section \ref{sec2}, we briefly review the equivalence of massive Thirring model and sine-Gordon model in 1+1 dimensional commutative spacetime. 
Section \ref{sec3} deals with the
twisted scalar and the twisted Dirac fields which are quantized in terms of the twisted creation and twisted annihilation operators. Section \ref{sec4} 
explores as to the noncommutative S-operator for both the twisted quantum field theories. In sections \ref{sec5} and \ref{sec6}, we analyze n-point correlation 
functions pertaining
to the noncommutative sine-Gordon model and massive Thirring model respectively. The twisted bosonization rules that establish equivalence between these two models are summarized in section \ref{sec7}. 
\section{Equivalence between sine-Gordon model and massive Thirring model in 1+1 dimensional commutative spacetime\label{sec2}}
Coleman \cite{Sid} showed that the Massive Thirring Model and the Sine-Gordon Model are equivalent to each other in the following sense: Any order correlation 
functions pertaining to both the models
turn out to have the same structure provided
specific identifications of the coupling constants and mass parameters of the two models are made.
The sine-Gordon model (SG) is a renormalizable field theory of a single scalar boson
field $\phi$ in 1+1 dimensions. The Lagrangian for SG model reads:
\begin{equation} \mathcal{L_{SG}}= \frac{1}{2} \partial_{\mu}\phi\partial^{\mu}\phi + \frac{\alpha_0}{\beta^2}\cos\beta\phi+\gamma_0\end{equation}
where $\alpha_0$, $\beta$ and $\gamma_0$ are real parameters. In order to establish the equivalence between the two models,
 we shall display their n-point functions. We define:
\begin{equation}A_{\pm}(x) = :e^{\pm i\beta\phi(x)}:\end{equation}
where :: is the normal-ordering operation defined by the mass $m$ (see \cite{Sid}).
Now the free field vacuum expectation value of $\prod_{i=1}^{n}A_{+}(x_i)A_{-}(y_i)$ is evaluated as:
\begin{equation}
\left\langle 0\left|T\prod_{i=1}^{n}A_{+}(x_i)A_{-}(y_i)\right|0\right\rangle = \frac{\prod_{i>j}[(x_i-x_j)^2(y_i-y_j)^2C^2m^4]^{\frac{\beta^2}{4\pi}}}
 {\prod_{i,j} [(x_i-y_i)^2Cm^2]^{\frac{\beta^2}{4\pi}}} \label{sg}
\end{equation}
Where $C$ is a numerical constant and $m$ is a mass parameter.
The massive Thirring model (MT) is a field theory of a single spin-1/2 fermion field with
a current-current interaction in 1+1 dimensions. The Lagrangian for MT model reads:
\begin{equation} \mathcal{L_{MT}} =\bar \psi(i\not\partial -m_0)\psi -\frac{1}{2}g j^{\mu}j_{\mu} \end{equation}
Now we define a renormalized scalar density:
\begin{equation} \sigma_{\pm} = \frac{1}{2}Z \bar\psi(x)(1\pm \gamma^5)\psi(x)\end{equation}
where $Z$ is the cutoff dependent multiplicative renormalization constant.
The vacuum expectation value of $\prod_{i=1}^{n}\sigma_{+}(x_i)\sigma_{-}(y_i)$ using Kaliber's formula \cite{kal} yields:
\begin{equation}
\left\langle 0\left|T\prod_{i=1}^{n}\sigma_{+}(x_i)\sigma_{-}(y_i)\right|0\right\rangle = \frac{\prod_{i>j}[(x_i-x_j)^2(y_i-y_j)^2 M^4]^{\frac{1}{1+g/\pi}}}
 {\prod_{i,j} [(x_i-y_i)^2 M^2]^{\frac{1}{1+g/\pi}}}\label{mt}
\end{equation}
where $g$ is a coupling constant and $M$ a mass parameter which absorbs all renormalization constants and numerical factors.
Comparison of equations (\ref{sg}) and (\ref{mt}) leads to the fact that the two perturbative field theories are
identical provided the following identifications are made:
\begin{eqnarray}
 \sigma_{\pm} &=& \frac{1}{2}A_{\pm} \Rightarrow \frac{1}{2}Z \bar\psi(x)(1\pm \gamma^5)\psi(x) = \frac{1}{2}
 :e^{\pm i\beta\phi(x)}:\nonumber\\
M^2 &=& Cm^2\nonumber\\
\frac{1}{1+g/\pi} &=& \frac{\beta^2}{4\pi}
\end{eqnarray}
Thus the two field theoretic models: sine-Gordon and massive Thirring are equivalent to each other as long as their perturbation expansions converge.
\section{Noncommutative quantum field theories in 1+1 dimensional spacetime \label{sec3}}
The effect of spacetime noncommutativity in QFT is
incorporated using the Moyal star product. Given a local QFT on a commutative
space-time, it can be generalized to a noncommutative space-time
which amounts to replacing ordinary local product by a Moyal star product. The Moyal star product is a mapping from operators (say $\hat A(\hat x)$, $\hat B(\hat x)$)
to functions ($A( x)$, $B( x)$) defined by:

\begin{equation}
\hat A(\hat x)\hat B(\hat x) \Leftrightarrow  (A*B)(x) = e^{\frac{i}{2}\theta^{\mu\nu}\partial_{\mu}^x\partial_{\nu}^y}A(x)B(y)|_{y=x}
\end{equation}

Quantum field theory on noncommutative spacetime with $\theta^{0i}\ne0$ is shown to be manifestly unitary \cite{un1}. Unitary quantum mechanics with 
time-space noncommutativity
has been developed in \cite{un2}.

\subsection{Twisted scalar field}
Consider a free scalar field
$$\phi(x) = \int{\frac{dk^1}{\sqrt{2\pi}2k^0}[a_k^{\theta}e^{-ik.x}+a_k^{\theta \dagger}e^{ik.x}]},~~~~~k^0=\sqrt{{k^1}^2 + m^2} $$
quantized via twisted commutation relations \cite{sac1} given as:
\begin{eqnarray}
a_p^{\theta} a_k^{\theta} &=& e^{ip\wedge k} a_k^{\theta} a_p^{\theta},\nonumber\\
a_p^{\theta} a_k^{\theta\dagger} &=& e^{-ip\wedge k} a_k^{\theta\dagger} a_p^{\theta} + 2p_0\delta(\vec p-\vec k)
\end{eqnarray}
The twisted creation and annihilation operators $a_p^{\theta}$ and  $a_p^{\theta\dagger}$ are related to the untwisted ones $a_p^{\dagger}$ and $a_p$ via`` dressing
 transformations'' [which were first discussed in \cite{dr1,dr2,dr3}]:
\begin{equation}
 a_p^{\theta}  = a_pe^{-\frac{i}{2}p\wedge P},~~~
a_p^{\theta\dagger}  = a_p^{\dagger}e^{\frac{i}{2}p\wedge P}.
\end{equation}
Thus twisted creation and annihilation operators $a_p^{\theta}$ and  $a_p^{\theta\dagger}$ act on the same Fock space as that of the
 untwisted ones $a_p^{\dagger}$ and  $a_p$.
The momentum operator of the scalar field is given by:
$$P^{\mu} = \int{\frac{dp^1}{2\pi2p^0}p^{\mu}a_p^{\theta\dagger}a_p^{\theta}} = \int{\frac{dp^1}{2\pi2p^0}p^{\mu}a_p^{\dagger}a_p};~~~[P^{\mu}, \phi] = -i\partial_{\mu}\phi$$
The twisted and untwisted n-particle states are related to each other as follows:
\begin{equation}
|p_1, p_2, ...p_n\rangle_{\theta} = a_{p_{1}}^{\theta\dagger}a_{p_{2}}^{\theta\dagger} .
...a_{p_{n}}^{\theta\dagger}|0\rangle = e^{\frac{i}{2}\sum_{i<j}p_i\wedge p_j}|p_1, p_2, ...p_n\rangle \label{bs}
\end{equation}
The commutator of $\phi(x)$ and $\phi(y)$ turns out:
\begin{eqnarray}
[\phi(x), \phi(y)] &=&
\int{\frac{dp^1dk^1}{2\pi2p_02k_0}}\left[e^{-i(p.x+k.y)}(e^{-ip\wedge k}-1)a_k^{\theta}a_p^{\theta}\right. \nonumber\\
&+&e^{i(p.x+k.y)}(e^{ip\wedge
 k}-1)a_k^{\theta\dagger}a_p^{\theta\dagger}\nonumber\\
&+&e^{-i(p.x-k.y)}\{(e^{ip\wedge
 k}-1)a_k^{\theta\dagger}a_p^{\theta}+2p_0\delta(\vec p-\vec
 k)\}\nonumber\\
 &+&\left.e^{i(p.x-k.y)}\{(1-e^{-ip\wedge
 k})a_p^{\theta\dagger}a_k^{\theta}-2p_0\delta(\vec p-\vec k)\}\right]
\end{eqnarray}
The twisted scalar fields at any two spacetime points with
space-like separation do not commute, however, $\langle
0|[\phi(x), \phi(y)]|0\rangle$ is zero for space-like separation.
\subsection{Twisted Dirac field}
A free Dirac field can be Fourier decomposed as
$$\psi(x) = \int{\frac{dk^1}{\sqrt{2\pi}2k^0}[b_k^{\theta}u(k)e^{-ik.x}+d_k^{\theta \dagger}v(k)e^{ik.x}]},~~~~~k^0=\sqrt{{k^1}^2 + m^2} $$
where $b_k^{\theta}$ and $d_k^{\theta \dagger}$ obey the following twisted anticommutation relations \cite{sac1,sac2} given as:
\begin{eqnarray}
b_p^{\theta} b_k^{\theta} &=&- e^{ip\wedge k} b_k^{\theta} b_p^{\theta},~~~
b_p^{\theta} b_k^{\theta\dagger} =- e^{-ip\wedge k} b_k^{\theta\dagger} b_p^{\theta} + 2p_0\delta(\vec p-\vec k),\nonumber\\
d_p^{\theta} d_k^{\theta} &=&- e^{ip\wedge k} d_k^{\theta} d_p^{\theta},~~~
d_p^{\theta} d_k^{\theta\dagger} =- e^{-ip\wedge k} d_k^{\theta\dagger} d_p^{\theta} + 2p_0\delta(\vec p-\vec k)
\end{eqnarray}
The twisted creation and annihilation operators $b_p^{\theta\dagger},d_p^{\theta\dagger}$ and $b_p^{\theta}, d_p^{\theta}$  are
related to the untwisted ones $b_p^{\dagger},d_p^{\dagger}$ and  $b_p, d_p$ as:
\begin{eqnarray}
 b_p^{\theta}  &=& b_pe^{-\frac{i}{2}p\wedge P} ; d_p^{\theta}  = d_pe^{-\frac{i}{2}p\wedge P},\nonumber\\
b_p^{\theta\dagger}  &=& b_p^{\dagger}e^{\frac{i}{2}p\wedge P} ; d_p^{\theta\dagger}  = d_p^{\dagger}e^{\frac{i}{2}p\wedge P}
\end{eqnarray}
The twisted creation and annihilation operators act on the same Fock space as that of their untwisted counterparts.
The total momentum operator of the Dirac field can be written as:
$$P^{\mu} = \int{\frac{dp^1}{2\pi2p^0}p^{\mu}[ b_p^{\theta\dagger}b_p^{\theta}} + d_p^{\theta\dagger}d_p^{\theta} ]=
\int{\frac{dp^1}{2\pi2p^0}p^{\mu}[ b_p^{\dagger}b_p} + d_p^{\dagger}d_p ];~~~[P^{\mu}, \psi] = -i\partial_{\mu}\psi$$
The twisted and untwisted n-particle fermionic states are related to each other as:
$$|p_1,s_1; p_2,s_2; ...p_n,s_n\rangle_{\theta} = b_{p_{1}}^{s_1\theta\dagger}b_{p_{2}}^{s_2\theta\dagger} ....
b_{p_{n}}^{s_n\theta\dagger}|0\rangle = e^{\frac{i}{2}\sum_{i<j}p_i\wedge p_j}|p_1,s_1; p_2,s_2; ...p_n,s_n\rangle $$
$$|p_1,s_1; p_2,s_2; ...p_n,s_n\rangle_{\theta} = d_{p_{1}}^{s_1\theta\dagger}d_{p_{2}}^{s_2\theta\dagger} ....
d_{p_{n}}^{s_n\theta\dagger}|0\rangle =
e^{\frac{i}{2}\sum_{i<j}p_i\wedge p_j}|p_1,s_1; p_2,s_2;
...p_n,s_n\rangle $$ The anticommutator of the twisted Dirac fields
$\psi(x)$ and $\bar\psi(y)$ is:
\begin{eqnarray}
\{\psi(x), \bar\psi(y)\} &=&
\int{\frac{dp^1dk^1}{2\pi2p_02k_0}}\left[e^{-i(p.x-k.y)}\{(1-e^{-ip\wedge
k})
b_k^{\theta\dagger}b_p^{\theta}+2p^0\delta(\vec p-\vec k)\}u(p)\bar u(k)\right.\nonumber\\
&+&e^{-i(p.x+k.y)}(1-e^{-ip\wedge
 k})d_k^{\theta}b_p^{\theta}u(p)\bar v(k)+e^{i(p.x+k.y)}(1-e^{-ip\wedge
 k})b_k^{\theta\dagger}d_p^{\theta\dagger}v(p)\bar u(k)\nonumber\\
 &+&\left.e^{i(p.x-k.y)}\{(1-e^{-ip\wedge
 k})d_p^{\theta\dagger}d_k^{\theta}+2p_0\delta(\vec p-\vec k)\}v(p)\bar v(k)\right]
\end{eqnarray}
The twisted Dirac fields at any two spacetime points with
space-like separation do not anticommute, however, $\langle
0|\{\psi(x), \bar\psi(y)\}|0\rangle$ is zero for space-like
separation.
\section{Noncommutative S-operator \label{sec4}}
The S-operator $ S_{\theta}$ for twisted quantum field theory reads:
\begin{equation} S_{\theta} = \mathcal{T}e^{-i\int_{-\infty}^{+\infty}H_I^{\theta}dx}\end{equation}
In order to study scattering theory on noncommutative spacetime, we shall analyze the S-operator pertaining to the following interaction
 Hamiltonians of noncommutative QFT's:
\begin{eqnarray}
 H_{I}^{\theta}& =&\frac{\lambda}{4!}\phi_*^n(x)= \frac{\lambda}{4!}\phi(x)*\phi(x)*\phi(x)*......*\phi(x) \\
 H_{I}^{'\theta} &= &  g\bar{\psi}(x)*\gamma^{\mu}\psi(x)*\bar{\psi}(x)*\gamma_{\mu}\psi(x).
\end{eqnarray}
 Let us consider the leading order terms corresponding to
$H_{I}^{\theta}$ and $H_{I}^{'\theta}$:
\begin{eqnarray}
 S_{\theta }^{(1)} &=& \frac{-i\lambda}{4!}\int{d^2x} (\phi*\phi*...*\phi)(x)\\
 S_{\theta }^{'(1)} &=& -ig\int{d^2x} (\bar{\psi}*\gamma^{\mu}\psi*\bar{\psi}*\gamma_{\mu}\psi)(x)
\end{eqnarray}
We shall now focus on a typical term arising from the Fourier decomposition of field $\phi$ in $\hat S_{\theta }^{(1)} $ as:
\begin{eqnarray}
&& \frac{-i\lambda}{4!}\int{d^2x} a_{p_1}^{\theta}a_{p_2}^{\theta\dagger}.....a_{p_{n-1}}^{\theta}a_{p_{n}}^{\theta\dagger}
e^{-ip_1.x}*e^{ip_2.x}*.....*e^{-ip_{n-1}.x}*e^{ip_n.x}
\nonumber\\
&& = \frac{-i\lambda}{4!}\int{d^2x} a_{p_1}a_{p_2}^{\dagger}.....a_{p_{n-1}}a_{p_{n}}^{\dagger}e^{\frac{i}{2} 
(\sum_i(-1)^ip_i)\wedge P}e^{\frac{i}{2} \sum_{i<j}(-1)^{i+j}p_i\wedge p_j} e^{(\sum_{i=1}^{n}(-1)^ip_i.)x}\nonumber\\
&&\times~e^{-\frac{i}{2} \sum_{i<j}(-1)^{i+j}p_i\wedge p_j }\nonumber \\
&& = \frac{-i\lambda}{4!}\int{d^2x} a_{p_1}a_{p_2}^{\dagger}.....a_{p_{n-1}}a_{p_{n}}^{\dagger}e^{\frac{i}{2} (\sum_i(-1)^ip_i)\wedge P} e^{(\sum_{i=1}^{n}(-1)^ip_i).x}\nonumber\\
&&= \frac{-i\lambda}{4!}\int{d^2x} a_{p_1}a_{p_2}^{\dagger}.....a_{p_{n-1}}a_{p_{n}}^{\dagger}e^{(\sum_{i=1}^{n}(-1)^ip_i).x}
\end{eqnarray}
The last step could be obtained either by using
$$\int{d^2x} e^{(\sum_{i=1}^{n}(-1)^ip_i).x} = \delta(\sum_{i=1}^{n}(-1)^ip_i )$$
which converts $e^{\frac{i}{2} (\sum_i(-1)^ip_i)\wedge P}$ to unity, or using
$$\frac{-i\lambda}{4!}\int{d^2x} a_{p_1}a_{p_2}^{\dagger}.....a_{p_{n-1}}a_{p_{n}}^{\dagger}e^{(\sum_{i=1}^{n}(-1)^ip_i).x}  e^{\frac{i}{2} \overleftarrow{\partial}\wedge P}$$
which may be expressed as a sum of the corresponding commutative counterpart and a surface terms [which might be discarded]. In fact such correspondence of the noncommutative
S-operator with the commutative S-operator is true for all order \cite{sac3}.\vspace{.1cm}

We now explore the specific term stemming from  $\hat S_{\theta }^{'(1)} $ as:
\begin{eqnarray}
&& -ig\int{d^2x} b_{p_1}^{\theta\dagger}b_{p_2}^{\theta}b_{p_3}^{\theta\dagger}b_{p_{4}}^{\theta} e^{ip_1.x}*e^{-ip_2.x}*e^{ip_3.x}*e^{-ip_4.x}\nonumber\\
&& = -ig\int{d^2x} b_{p_1}^{\dagger}b_{p_2}b_{p_3}^{\dagger}b_{p_{4}}e^{\frac{i}{2} (p_1-p_2+p_3-p_4)
\wedge P}e^{-\frac{i}{2}[p_1\wedge(p_2-p_3-p_4)+p_2\wedge (p_3-p_4)+ p_3\wedge p_4 ]} \nonumber\\
&& \times~e^{i(p_1-p_2+p_3-p_4).x}e^{\frac{i}{2}[p_1\wedge(p_2-p_3-p_4)+p_2\wedge (p_3-p_4)+ p_3\wedge p_4 ]}\nonumber\\
&& = -ig\int{d^2x} b_{p_1}^{\dagger}b_{p_2}b_{p_3}^{\dagger}b_{p_{4}}e^{\frac{i}{2} (p_1-p_2+p_3-p_4)\wedge P}e^{i(p_1-p_2+p_3-p_4).x}\nonumber\\
&& = -ig\int{d^2x} b_{p_1}^{\dagger}b_{p_2}b_{p_3}^{\dagger}b_{p_{4}}e^{i(p_1-p_2+p_3-p_4).x}e^{\frac{i}{2} \overleftarrow{\partial}\wedge P}\nonumber\\
&& = -ig\int{d^2x} b_{p_1}^{\dagger}b_{p_2}b_{p_3}^{\dagger}b_{p_{4}}e^{i(p_1-p_2+p_3-p_4).x}
\end{eqnarray}
where in the last step we have discarded the surface terms. Thus to the leading order: 
$\hat S_{\theta }^{'(1)} = \hat S_0^{'(1)}$. This might be shown to be true for all orders.\vspace{0.1cm}

The S-matrix element ($S_{\theta}[p_4,p_3;p_1,p_2]$) of a noncommutative field theory, for instance, for the process: $\phi(p_1) + \phi(p_2) \to \phi(p_3) + \phi(p_4)$ could be expressed 
in terms of its commutative counterpart ($S_0[p_4,p_3;p_1,p_2]$) using (\ref{bs}) as:
\begin{equation}S_{\theta}[p_4,p_3;p_1,p_2] = e^{\frac{i}{2}[p_1\wedge p_2-p_3\wedge p_4]}S_0[p_4,p_3;p_1,p_2] \end{equation}
The phase factor stems solely from the incoming and outgoing multi-particle twisted states as $\hat S_{\theta} = \hat S_0$.
At this juncture it is worthy to note that noncommutative field theories [involving scalar field or/and matter field alone] are as much 
renormalizable as their commutative counterpart
since divergences in the S-matrix elements of the former springs solely from the S-matrix elements of the latter. 
\section{Noncommutative sine-Gordon model \label{sec5}}
The Lagrangian density for the noncommutative sine-Gordon model (NCSG) reads:
\begin{equation} \mathcal{L_{NCSG}}= \frac{1}{2} \partial_{\mu}\phi*\partial^{\mu}\phi + \frac{\alpha_0}{\beta^2}\cos_*\beta\phi+\gamma_0\end{equation}
where $\cos_*\beta\phi \equiv \frac{1}{2}(e_*^{i\beta\phi}+e_*^{-i\beta\phi}) \equiv 1-\frac{\beta^2}{2!}\phi*\phi + \frac{\beta^4}{4!}\phi*\phi*\phi*\phi+.....$.\\
The Euler-Lagrange equation [after rescaling $\beta\phi \to \phi$] gives:
\begin{equation}
  (\frac{\partial^2}{\partial t^2}-  \frac{\partial^2}{\partial x^2})\phi +\alpha_0\sin_*\phi= 0\label{el}
\end{equation}
The solitonic solution of the classical field equation of the sine-Gordon model on commutative spacetime is known to be:
\begin{equation}
 \phi(x,t) = 4tan^{-1}[e^{\frac{\sqrt{\alpha_0}(x-vt)}{\sqrt{1-v^2}}}]
\end{equation}
which turns out to be the solution of equation (\ref{el}) as well. To see this, let us now consider 
the following term appearing in the Euler-Lagrange equation (\ref{el}):
\begin{eqnarray}
 \sin_*\phi &= &\phi -\frac{ \phi_*^3}{3!} + \frac{ \phi_*^5}{5!}-.....\nonumber\\
 &=& \phi -4tan^{-1}[e^{\frac{\sqrt{\alpha_0}( x-vt)}{\sqrt{1-v^2}}}]*4tan^{-1}[e^{\frac{\sqrt{\alpha_0}( x-vt)}
 {\sqrt{1-v^2}}}]*4tan^{-1}[e^{\frac{\sqrt{\alpha_0}( x-vt)}{\sqrt{1-v^2}}}]+...\label{si}
\end{eqnarray}
The general form corresponding to the second term on RHS in the above equation (\ref{si}) can  be simplified to:
\begin{equation}
e^{n_1\frac{\sqrt{\alpha_0}(x-vt)}{\sqrt{1-v^2}}}*e^{n_2\frac{\sqrt{\alpha_0}(x-vt)}{\sqrt{1-v^2}}}*e^{n_3\frac{\sqrt{\alpha_0}(x-vt)}{\sqrt{1-v^2}}}
= e^{(n_1+n_2+n_3)\frac{\sqrt{\alpha_0}(x-vt)}{\sqrt{1-v^2}}}
\end{equation}
where $n_1, n_2$ and $n_3$ are positive integers. In fact, we can have
\begin{equation}
 \sin_*\phi = \sin\phi
\end{equation}
Thus the solution of classical field equation $\phi(x,t)$ on commutative spacetime happens to be also the solution of the corresponding noncommutative theory.
\subsection{n-point correlation function in NCSG model}
We are now ready to establish the correspondence between the NCSG and the NCMT models, we evaluate the following n-point function:
$$\left\langle 0\left|T\prod_{i=1}^{n}:e_*^{i\beta\phi(x_i)}::e_*^{i\beta\phi(y_i)}:\right|0\right\rangle$$
where, $ e_*^{i\beta\phi(x)} \equiv 1 + i\beta\phi + \frac{(i\beta)^2}{2!}\phi*\phi + \frac{(i\beta)^3}{3!}\phi*\phi*\phi ..... $\\
Now,
\begin{eqnarray}
&&\left\langle 0\left|T\prod_{i=1}^{n}:e_*^{i\beta\phi(x_i)}::e_*^{i\beta\phi(y_i)}:\right|0\right\rangle \nonumber \\
&&= 
\left\langle 0\left|T:e_*^{i\beta\phi(x_1)}::e_*^{i\beta\phi(y_1)}:...:e_*^{i\beta\phi(x_n)}::e_*^{i\beta\phi(y_n)}:\right|0\right\rangle\nonumber\\
&&= \left\langle 0\left|T:\sum_{n_1=0}^{\infty}\frac{({i\beta\phi_*(x_1)})^{n_1}}{n_1!}::\sum_{m_1=0}^{\infty}
\frac{({i\beta\phi_*(y_1)})^{m_1}}{m_1!}:.....\right.\right.\nonumber\\
&&\times \left.\left.:\sum_{n_n=0}^{\infty}\frac{({i\beta\phi_*(x_n)})^{n_n}}{n_n!}::\sum_{m_n=0}^{\infty}\frac{({i\beta\phi_*(y_n)})^{m_n}}{m_n!}:\right|0\right\rangle 
\end{eqnarray}
For our purpose, we shall consider the specific non-vanishing term (containing equal number of creation and annihilation operators)
arising from:
\begin{equation}  \left\langle 0\left|T:\phi_*^n(x_1)::\phi_*^n(y_1):...
:\phi_*^n(x_n)::\phi_*^n(y_n):\right|0\right\rangle \label{npoint}\end{equation}
We shall use the following notations:
\begin{equation}
 \phi(x_1) = \int{\frac{dp_{11}}{(2\pi\sqrt{2\omega_{p_{11}}}}}[a_{p_{11}}e^{-ip_{11}.x_1} +a_{p_{11}}^{\dagger}e^{ip_{11}.x_1}]
\end{equation}
so that
\begin{equation}
 \phi(x_1)\phi(x_1) = \int{\frac{dp_{11}dp_{12}}{(2\pi)^2\sqrt{2\omega_{p_{11}}2\omega_{p_{12}}}}}[a_{p_{11}}e^{-ip_{11}.x_1} +a_{p_{11}}^{\dagger}e^{ip_{11}.x_1}]
 [a_{p_{12}}e^{-ip_{12}.x_1} +a_{p_{12}}^{\dagger}e^{ip_{12}.x_1}]
\end{equation}

Now the specific non-vanishing term of the n-point function\footnote{Here for nonzero n-point function, n has to be even to have equal numbers of
creation and annihilation operators. However we can always choose even number of field operators for nonzero n(either even or odd)-point function. 
It has been checked that the
conclusions remain unaltered for either even or odd n.}
\begin{eqnarray*}
&& \left \langle 0\left| {a_{p_{11}}^{\theta}...a_{p_{n1}}^{\theta} e^{-ip_{11}.x_1}*...*e^{-ip_{n1}.x_1}}
{a_{k_{11}}^{\theta}...a_{k_{n1}}^{\theta} e^{-ik_{11}.y_1}*...*e^{-ik_{n1}.y_1}}\right.\right.\\
&&\times.....{a_{p_{1(n/2)}}^{\theta}...a_{p_{n(n/2)}}^{\theta} e^{-ip_{1(n/2)}.x_{n/2}}*...*e^{-ip_{n(n/2)}.x_{n/2}}}\\
&&\times{a_{k_{1(n/2)}}^{\theta}... a_{k_{n(n/2)}}^{\theta} e^{-ik_{1(n/2)}.y_{n/2}}*...*e^{-ik_{n(n/2)}.y_{n/2}}}\\
&&\times{a_{p_{1(n/2+1)}}^{\theta\dagger}...a_{p_{n(n/2+1)}}^{\theta\dagger} e^{ip_{1(n/2+1)}.x_{n/2+1}}*...*e^{ip_{n (n/2+1)}.x_{n/2+1}}}\\
&&\times{a_{k_{1(n/2+1)}}^{\theta\dagger}... a_{k_{n(n/2+1)}}^{\theta\dagger} e^{ik_{1(n/2+1)}.y_{n/2+1}}*...*e^{ik_{n(n/2+1)}.y_{n/2+1}}}.....\\
&&\left.\left.\times{a_{p_{1 n}}^{\theta\dagger}...a_{p_{n n}}^{\theta\dagger} e^{ip_{1 n}.x_n}*...*e^{ip_{n n}.x_n}}
{a_{k_{1 n}}^{\theta\dagger}... a_{k_{n n}}^{\theta\dagger} e^{ik_{1 n}.y_n}*...*e^{ik_{n n}.y_n}}\right|0\right\rangle
\end{eqnarray*}
\begin{eqnarray*}
&&= \left \langle 0\left| {a_{p_{11}}...a_{p_{n1}} e^{-\frac{i}{2} (\sum_ip_{i1})\wedge P} e^{\frac{i}{2}\sum_{i<j}p_{i1}\wedge p_{j1}}
 e^{-i\sum_ip_{i1}.x_1}}  e^{-\frac{i}{2}\sum_{i<j}p_{i1}\wedge p_{j1}}\right.\right. \\
&&\times {a_{k_{11}}...a_{k_{n1}}}e^{-\frac{i}{2} (\sum_ik_{i1})\wedge P} e^{\frac{i}{2}\sum_{i<j}k_{i1}\wedge k_{j1}}
 e^{-i\sum_ik_{i1}.y_1}  e^{-\frac{i}{2}\sum_{i<j}k_{i1}\wedge k_{j1}}.....\\
&&\times{ {a_{p_{1n}}^{\dagger}a_{p_{2n}}^{\dagger}...a_{p_{nn}}^{\dagger} }}
e^{\frac{i}{2} (\sum_ik_{in})\wedge P} e^{\frac{i}{2}\sum_{i<j}p_{in}\wedge p_{jn}}
 e^{i\sum_ip_{in}.y_n}  e^{-\frac{i}{2}\sum_{i<j}p_{in}\wedge p_{jn}}\\
&&\left.\left.\times ~{ {a_{k_{1n}}^{\dagger}a_{k_{2n}}^{\dagger}...a_{k_{nn}}^{\dagger} }}
e^{\frac{i}{2} (\sum_ik_{in})\wedge P} e^{\frac{i}{2}\sum_{i<j}k_{in}\wedge k_{jn}}
 e^{i\sum_ik_{in}.y_n}  e^{-\frac{i}{2}\sum_{i<j}k_{in}\wedge k_{jn}}\right|0\right\rangle\\
&&= \left \langle 0\left| {a_{p_{11}}...a_{p_{n1}}
 e^{-i\sum_ip_{i1}.x_1}}{a_{k_{11}}...a_{k_{n1}}}
 e^{-i\sum_ik_{i1}.y_1} .....\right.\right. \\
&&\times ~{ { a_{p_{1n}}^{\dagger}...a_{p_{nn}}^{\dagger} }}
 e^{i\sum_ip_{in}.x_n} a_{k_{1n}}^{\dagger}...a_{k_{nn}}^{\dagger}
 e^{i\sum_ik_{in}.y_n} \\
&&\left.\left.\times e^{-\frac{i}{2}(\sum_{i}[p_{i1}+...+p_{i(n/2)}+ k_{i1}+...+k_{i(n/2)}])\wedge (\sum_{i}[p_{i(n/2+1)}+...+p_{in}+ k_{i(n/2+1)}+...+k_{in}])}\right|0\right\rangle
\end{eqnarray*}
 The following expression
\begin{eqnarray*}
 &&= \left \langle 0\left| {a_{p_{11}}a_{p_{21}}...a_{p_{n1}}
 }{a_{k_{11}}a_{k_{21}}... a_{k_{(n-1) 1}}a_{k_{n1}}}
 \right.\right. \\
&&\left.\left.....{a_{p_{1(n-1)}}^{\dagger}a_{p_{2(n-1)}}^{\dagger}...a_{p_{n(n-1)}}^{\dagger} }... { {a_{k_{1n}}^{\dagger}a_{k_{2n}}^{\dagger}...a_{k_{nn}}^{\dagger} }}
 \right|0\right\rangle
\end{eqnarray*}
 turns out to be the sum of the various terms each of which contains product of several Dirac delta functions. A typical term looks like
\begin{eqnarray}
\delta(k_{n(n/2)}-p_{1(n/2+1)})\delta(k_{(n-1)(n/2)}-p_{2(n/2+1)})......\delta(p_{21}-k_{(n-1)n})\delta(p_{11 }-k_{nn})
\end{eqnarray}
It is evident that the domain of support corresponding to each term pertaining to the product of several Dirac delta functions leads
 to
$$ p_{i1}=k_{(n+1-i) n},~ p_{i2}= k_{(n+1-i)(n-1)},...........,p_{in}=k_{(n+1-i)1};~~i=1,2,....,n$$
 Therefore $e^{-\frac{i}{2}(\sum_{i}[p_{i1}+...+p_{i(n/2)}+ k_{i1}+...+k_{i(n/2)}])\wedge (\sum_{i}[p_{i(n/2+1)}+...+p_{in}+ k_{i(n/2+1)}+...+k_{in}])}$ actually becomes one. Thus
\begin{eqnarray}
&&= \left \langle 0\left| {a_{p_{11}}...a_{p_{n1}}
 e^{-i\sum_ip_{i1}.x_1}}{a_{k_{11}}...a_{k_{n1}}}
 e^{-i\sum_ik_{i1}.y_1}..... \right.\right. \nonumber\\
&&\times ~{ { a_{p_{1n}}^{\dagger}...a_{p_{nn}}^{\dagger} }}
 e^{i\sum_ip_{in}.x_n} a_{k_{1n}}^{\dagger}...a_{k_{nn}}^{\dagger}
 e^{i\sum_ik_{in}.y_n} \nonumber\\
&&\left.\left.\times e^{-\frac{i}{2}(\sum_{i}[p_{i1}+...+p_{i(n/2)}+ k_{i1}+...+k_{i(n/2)}])\wedge (\sum_{i}[p_{i(n/2+1)}+...+p_{in}+ 
k_{i(n/2+1)}+...+k_{in}])}\right|0\right\rangle \nonumber\\
&&= \left \langle 0\left| {a_{p_{11}}...a_{p_{n1}}
 e^{-i\sum_ip_{i1}.x_1}}{a_{k_{11}}...a_{k_{n1}}}
 e^{-i\sum_ik_{i1}.y_1} .....\right.\right. \nonumber\\
&&\left.\left.\times ~{ { a_{p_{1n}}^{\dagger}...a_{p_{nn}}^{\dagger} }}
 e^{i\sum_ip_{in}.x_n} a_{k_{1n}}^{\dagger}...a_{k_{nn}}^{\dagger}
 e^{i\sum_ik_{in}.y_n} \right|0\right\rangle
\end{eqnarray}
Such correspondence between the noncommutative and commutative matrix elements might be extended to all non vanishing contributions of the n-point functions. Therefore
\begin{eqnarray}
\left\langle 0\left|T\prod_{i=1}^{n}:e_*^{i\beta\phi(x_i)}::e_*^{i\beta\phi(y_i)}:\right|0\right\rangle
 &= &\left\langle 0\left|T\prod_{i=1}^{n}:e^{i\beta\phi(x_i)}::e^{i\beta\phi(y_i)}:\right|0\right\rangle\nonumber\\
&=& \frac{\prod_{i>j}[(x_i-x_j)^2(y_i-y_j)^2C^2m^4]^{\frac{\beta^2}{4\pi}}}
 {\prod_{i,j} [(x_i-y_i)^2Cm^2]^{\frac{\beta^2}{4\pi}}}\label{ncsg}
\end{eqnarray}
\section{Noncommutative massive Thirring model \label{sec6}}
The  noncommutative massive Thirring model (NCMT) is described by the following Lagrangian density:
\begin{equation} \mathcal{L_{NCMT}} =\bar \psi(x)*(i\not\partial -m_0)\psi(x) -\frac{1}{2} gj_*^{\mu}*j_{*\mu} \end{equation}
Now we shall be interested in the calculation of the following quantity:
$$ \left \langle 0\left| T\prod_{i=1}^{n}\frac{1}{2}Z \bar\psi(x_i)*(1+ \gamma^5)\psi(x_i) \frac{1}{2}Z \bar\psi(y_i)*(1- \gamma^5)\psi(y_i)\right|0\right\rangle$$
A typical non vanishing term of the above expression is:
\begin{eqnarray*}
&& \left \langle 0\left| d_{p_1}^{\theta}b_{p_2}^{\theta}d_{k_1}^{\theta}b_{k_2}^{\theta}...d_{p_{n-1}}^{\theta}b_{p_{n}}^{\theta}d_{k_{n-1}}^{\theta}b_{k_n}^{\theta}\right.\right.\\
&&\times \prod_{j=1}^{n-1} [e^{-ip_{j}.x_i}*e^{-ip_{j+1}.x_i}e^{-ik_{j}.y_i}*e^{-ik_{j+1}.y_i}]\\
&& \times d_{p_{n+1}}^{\theta\dagger}b_{p_{n+2}}^{\theta\dagger}b_{k_{n+1}}^{\theta\dagger}
d_{k_{n+2}}^{\theta\dagger}...b_{p_{2(n-1)}}^{\theta\dagger}d_{p_{2n}}^{\theta\dagger}b_{k_{2(n-1)}}^{\theta\dagger}d_{k_{2n}}^{\theta\dagger}\\
&& \left.\left.\times \prod_{j=n+1}^{2(n-1)} [e^{ip_{j}.x_i}*e^{ip_{j+1}.x_i}e^{ik_{j}.y_i}*e^{ik_{j+1}.y_i}]
\right|0\right\rangle\\
&&= \left \langle 0\left| {d_{p_1}b_{p_2}...d_{k_{n-1}}b_{k_{n}}e^{\frac{i}{2} \sum_{j=1}^{n-1}[p_j\wedge p_{j+1}+k_j\wedge k_{j+1}] }}\right.\right.\\
&&\times e^{\frac{i}{2} \sum_{j=1}^{n}p_j\wedge\sum_{j=1}^{n} k_j }e^{-\frac{i}{2} \sum_{j=1}^{n}(p_j+k_j)\wedge P }b_{p_{n+1}}^{\dagger}
d_{p_{n+2}}^{\dagger}...b_{k_{2(n-1)}}^{\dagger}d_{k_{2n}}^{\dagger} \\
&& \times e^{\frac{i}{2} \sum_{j=n+1}^{2n-1}[p_j\wedge p_{j+1}+k_j\wedge k_{j+1}] }
 e^{\frac{i}{2} \sum_{j=n+1}^{2n}p_j\wedge\sum_{j=n+1}^{2n} k_j }e^{\frac{i}{2} \sum_{j=n+1}^{2n-1}(p_j+k_j)\wedge P }\\
&&  \times{ e^{-i\sum_{j=1}^{n-1}[(p_{j}+p_{j+1}).x_j+(k_j+k_{j+1}).y_j]}e^{-\frac{i}{2}\sum_{j=1}^{n-1}[p_j\wedge p_{j+1}+k_j\wedge k_{j+1}] }}\\
&&\left.\left.\times{ e^{i\sum_{j=n+1}^{2n-1}[(p_{j}+p_{j+1}).x_j+(k_j+k_{j+1}).y_j]}e^{-\frac{i}{2}\sum_{j=n+1}^{2n-1}[p_j\wedge p_{j+1}+k_j\wedge k_{j+1}] }}
\right|0\right\rangle\\
&&= \left \langle 0\left| {d_{p_1}b_{p_2}...d_{k_{n-1}}b_{k_{n}}
 e^{\frac{i}{2} \sum_{j=1}^{n}p_j\wedge\sum_{j=1}^{n} k_j } b_{p_{n+1}}^{\dagger}
d_{p_{n+2}}^{\dagger}...b_{k_{2(n-1)}}^{\dagger}d_{k_{2n}}^{\dagger} }\right.\right.\\
&& \times
 e^{\frac{i}{2} \sum_{j=n+1}^{2(n-1)}p_j\wedge\sum_{j=n+1}^{2n} k_j }e^{-\frac{i}{2}[\sum_{j=1}^{n}(p_j+ k_j)\wedge \sum_{j=n+1}^{2n}(p_j+k_j)] }\\
&& \left.\left. \times{ e^{-i\sum_{j=1}^{n-1}[(p_{j}+p_{j+1}).x_i+(k_j+k_{j+1}).y_j]}}
{ e^{i\sum_{j=n+1}^{2(n-1)}[(p_{j}+p_{j+1}).x_i+(k_j+k_{j+1}).y_j]}}
\right|0\right\rangle
\end{eqnarray*}
The object $\left \langle 0\left| {d_{p_1}b_{p_2}...d_{k_{n-1}}b_{k_{n}}  b_{p_{n+1}}^{\dagger}
d_{p_{n+2}}^{\dagger}...b_{k_{2(n-1)}}^{\dagger}d_{k_{2n}}^{\dagger}}\right|0\right\rangle$ is proportional to:
$$\delta(k_{n}-p_{n+1})\delta(k_{n-1}-p_{n+2})....\delta(k_{2n-1}-p_{2})\delta(k_{2n}-p_{1})$$
which leads to the constraints: $k_i = p_{2n+1-i}; i=1,2,...n$. It is
straightforward to notice that all the following exponential terms
$$e^{\frac{i}{2} \sum_{j=1}^{n}p_j\wedge\sum_{j=1}^{n} k_j } e^{\frac{i}{2} \sum_{j=n+1}^{2n-1}p_j
\wedge\sum_{j=n+1}^{2n} k_j }e^{-\frac{i}{2}[\sum_{j=1}^{n}(p_j+ k_j)\wedge \sum_{j=n+1}^{2n}(p_j+k_j)] }$$
could be set to unity over the domain of support of the above product of Dirac delta functions. Thus
\begin{eqnarray}
&&= \left \langle 0\left| {d_{p_1}b_{p_2}...d_{k_{n-1}}b_{k_{n}}
 e^{\frac{i}{2} \sum_{j=1}^{n}p_j\wedge\sum_{j=1}^{n} k_j } b_{p_{n+1}}^{\dagger}
d_{p_{n+2}}^{\dagger}...b_{k_{2(n-1)}}^{\dagger}d_{k_{2n}}^{\dagger} }\right.\right.\nonumber\\
&& \times
 e^{\frac{i}{2} \sum_{j=n+1}^{2(n-1)}p_j\wedge\sum_{j=n+1}^{2n} k_j }e^{-\frac{i}{2}[\sum_{j=1}^{n}(p_j+ k_j)\wedge\sum_{j=n+1}^{2n}(p_j+k_j)] }\nonumber\\
&& \left.\left. \times{ e^{-i\sum_{j=1}^{n-1}[(p_{j}+p_{j+1}).x_i+(k_j+k_{j+1}).y_j]}}
{ e^{i\sum_{j=n+1}^{2(n-1)}[(p_{j}+p_{j+1}).x_i+(k_j+k_{j+1}).y_j]}}
\right|0\right\rangle \nonumber\\
&&= \left \langle 0\left| {d_{p_1}b_{p_2}...d_{k_{n-1}}b_{k_{n}}
  b_{p_{n+1}}^{\dagger}
d_{p_{n+2}}^{\dagger}...b_{k_{2(n-1)}}^{\dagger}d_{k_{2n}}^{\dagger} }\right.\right.\nonumber\\
&& \left.\left. \times{ e^{-i\sum_{j=1}^{n-1}[(p_{j}+p_{j+1}).x_i+(k_j+k_{j+1}).y_j]}}
{ e^{i\sum_{j=n+1}^{2(n-1)}[(p_{j}+p_{j+1}).x_i+(k_j+k_{j+1}).y_j]}}
\right|0\right\rangle
\end{eqnarray}
This noncommutative-commutative equivalence extends to the all non vanishing terms of the n-point function. Thus
\begin{eqnarray}
&&\left \langle 0\left| T\prod_{i=1}^{n}\frac{1}{2} Z\bar\psi(x_i)*(1+ \gamma^5)\psi(x_i) \frac{1}{2}Z \bar\psi(y_i)*(1- \gamma^5)\psi(y_i)\right|0\right\rangle \nonumber\\
&&=\left \langle 0\left| T\prod_{i=1}^{n}\frac{1}{2}Z \bar\psi(x_i)(1+ \gamma^5)\psi(x_i) \frac{1}{2} Z\bar\psi(y_i)(1- \gamma^5)\psi(y_i)\right|0\right\rangle\nonumber\\
&&= \frac{\prod_{i>j}[(x_i-x_j)^2(y_i-y_j)^2 M^4]^{\frac{1}{1+g/\pi}}}
 {\prod_{i,j} [(x_i-y_i)^2 M^2]^{\frac{1}{1+g/\pi}}} \label{nctm}
\end{eqnarray}
\section{Equivalence between NCSG and NCMT models \label{sec7}}
The two perturbative noncommutative field theories, on comparing equations (\ref{ncsg}) and (\ref{nctm}), turn out to be identical 
provided we make the following identification:
\begin{eqnarray}
 \frac{1}{2}Z \bar\psi(x)*(1\pm \gamma^5)\psi(x)&=& \frac{1}{2}:e_*^{\pm i\beta\phi(x)}:\nonumber\\
M^2 &=& Cm^2\nonumber\\
\frac{1}{1+g/\pi} &=& \frac{\beta^2}{4\pi}
\end{eqnarray}
We shall now evaluate the vacuum expectation value of the commutator $[\partial_{\nu}\phi(x), e_*^{\pm
i\beta\phi(y)}]$.
\begin{eqnarray}\langle0|[\partial_{\nu}\phi(x),: e_*^{\pm
i\beta\phi(y)}:]|0\rangle& =& \langle 0|[\partial_{\nu}\phi(x), :e^{\pm
i\beta\phi(y)}:]|0\rangle \nonumber\\
& =& \pm\beta g_{\nu 0}\delta(\vec x-\vec y)\langle0|:e^{\pm
i\beta\phi}:|0\rangle\end{eqnarray}
The matrix element of $[\bar\psi(x)*\gamma_{\mu}\psi(x),\frac{1}{2}Z\bar\psi(y)*(1\pm\gamma_{5})\psi(y)]$ between the vacuum states reads:
\begin{eqnarray}
&&\langle0|[\bar\psi(x)*\gamma_{\mu}\psi(x),\frac{1}{2}Z\bar\psi(y)*(1\pm\gamma_{5})\psi(y)]|0\rangle \nonumber\\
&& =
 \langle 0|[\bar\psi(x)\gamma_{\mu}\psi(x),\frac{1}{2}Z\bar\psi(y)(1\pm\gamma_{5})\psi(y)]|0\rangle \nonumber\\
&& = \pm2(1+g/\pi)^{-1}\epsilon^{\mu\nu}\beta g_{\nu 0}\delta(\vec
x-\vec
y)\langle0|\frac{1}{2}Z\bar\psi(1\pm\gamma_{5})\psi|0\rangle\end{eqnarray}
From $\frac{1}{2}Z \bar\psi(x)*(1\pm \gamma^5)\psi(x)=
\frac{1}{2}:e_*^{\pm i\beta\phi(x)}:$, we can have
\begin{equation}\bar\psi(x)*\gamma^{\mu}\psi(x) = - \frac{g}{2\pi}\epsilon^{\mu\nu}\partial_{\nu}\phi(x)\end{equation}
which holds true for the matrix elements of the above commutators (see Appendix B). 
\section{Conclusions \label{sec8}}
We have extended the rules for bosonization in two dimensional spacetime to noncommutative spacetime with twisted quantization conditions. While the past research
concentrated on issues of classical integrability, we have focused on bosonization.
We have also checked using our bosonization rules, the duality between massive Thirring model and sine-Gordon model.\vspace{0.1cm}
 
 We have shown the static finite energy solution, solitons, of the classical field equation for commutative 1+1 sine-Gordon model is also the static 
 finite energy solution
of the classical field equation for its noncommutative counterpart. Therefore, the composite field versus fundamental field correspondence pertaining to
the two models persists even in
the noncommutative spacetime since the quantum soliton, quantized static solutions using semi-classical method, of sine-Gordon model could be recognized as 
the fundamental fermion
of the noncommutative massive Thirring model.\vspace{0.1cm}

 The sine-Gordon field in NC spacetime requires twisted bosonic
 rules. The dual massive Thirring model requires twisted fermionic quantization. Our bosonization rules are consistent with these requirements. \vspace{0.1cm}
 
 The bosonization in commutative spacetime is considered Abelian bosonization.
 Non-Abelian bosonization has further interesting structures \cite{wit}. This is anticipated to be even more interesting and subtle in 
 NC spacetime which will be explored later.
\section*{Appendix A}
\begin{eqnarray}
b^{\theta}(p) b^{\theta\dagger}(p') + e^{-p\wedge p'}b^{\theta\dagger}(p')b^{\theta}(p)
&=&  e^{i\pi\int_{p^1}^\infty\frac{dk^1}{2k^0}a^{\dagger}(k)a(k) }a^{\theta}(p)a^{\theta\dagger}(p')e^{-i\pi\int_{p^{'1}}^\infty\frac{dk^1}{2k^0}a^{\dagger}(k)a(k) }\nonumber\\
 &+& e^{-ip\wedge p'}a^{\theta\dagger}(p')
e^{-i\pi\int_{p^{'1}}^\infty\frac{dk^1}{2k^0}a^{\dagger}(k)a(k) }e^{i\pi\int_{p^1}^\infty\frac{dk^1}{2k^0}a^{\dagger}(k)a(k) }a^{\theta}(p);\nonumber\\
&=& e^{-ip\wedge p'} e^{i\pi\int_{p^1}^\infty\frac{dk^1}{2k^0}a^{\dagger}(k)a(k) }a^{\theta\dagger}(p')a^{\theta}(p)e^{-i\pi\int_{p^{'1}}^\infty\frac{dk^1}{2k^0}a^{\dagger}(k)a(k) }\nonumber\\
&+&e^{-i\pi\int_{p^1}^\infty\frac{dk^1}{2k^0}a^{\dagger}(k)a(k) }2p^0\delta(p-p')e^{-i\pi\int_{p^{'1}}^\infty\frac{dk^1}{2k^0}a^{\dagger}(k)a(k) }\nonumber\\
 &+& e^{-ip\wedge p'}a^{\theta\dagger}(p')
e^{-i\pi\int_{p^{'1}}^\infty\frac{dk^1}{2k^0}a^{\dagger}(k)a(k) }e^{i\pi\int_{p^1}^\infty\frac{dk^1}{2k^0}a^{\dagger}(k)a(k) }a^{\theta}(p);\nonumber\\
&=& e^{-ip\wedge p'} a^{\theta\dagger}(p')a^{\theta}(p)e^{i\pi\int_{p^1}^\infty\frac{dk^1}{2k^0}a^{\dagger}(k)a(k) }e^{-i\pi\int_{p^{'1}}^\infty\frac{dk^1}{2k^0}a^{\dagger}(k)a(k) }\nonumber\\
&+&2p^0\delta(p-p')\nonumber\\
 &+& e^{-ip\wedge p'}
a^{\theta\dagger}(p')
e^{i \pi }a^{\theta}(p)e^{-i \pi\int_{p^{'1}}^\infty\frac{dk^1}{2k^0}a^{\dagger}(k)a(k) }e^{i\pi\int_{p^1}^\infty\frac{dk^1}{2k^0}a^{\dagger}(k)a(k) }\nonumber\\
&=&2p^0\delta(p-p')
\end{eqnarray}
Where we have used:
\begin{eqnarray*}
&& a^{\theta}(p) a^{\theta\dagger}(p') = e^{-p\wedge p'}a^{\theta\dagger}(p')a^{\theta}(p) +  2p^0\delta(p-p');\\
&& e^{i\pi\int_{p^{'1}}^\infty dk^1\delta(k-p) }e^{-i\pi\int_{p^1}^\infty dk^1\delta(k-p) } = e^{i\pi\int_{p^{'1}}^{p^1} dk^1\delta(k-p) } = e^{i \pi } = -1
\end{eqnarray*}
\section*{Appendix B}
Let us evaluate the commutator: $[\partial_{\nu}\phi(x), :e_*^{\pm
i\beta\phi(y)}:]$. 
For the twisted field $\phi(x)$, we can have:
\begin{eqnarray}
\phi(x) &=&\phi_0(x)e^{\frac{1}{2} \overleftarrow{\partial} \wedge P}\nonumber\\
 e_*^{\pm
i\beta\phi(x)}& =& 1+i\beta\phi+\frac{(i\beta)^2}{2!}\phi(x)*\phi(x) +.....= e^{\pm i\beta\phi_0(x)}e^{\frac{1}{2} \overleftarrow{\partial} \wedge P}
\end{eqnarray}
Therefore the commutator could be expanded in a series in noncommutative parameter $\theta$,
$$[\partial_{\nu}\phi(x), e_*^{\pm
i\beta\phi(y)}] = [\partial_{\nu}\phi_0(x)e^{\frac{1}{2} \partial \wedge P}, e^{\pm i\beta\phi_0(y)}e^{\frac{1}{2} \partial \wedge P}] = a_0(x,y) + \theta a_1(x,y) + O(\theta^2)$$
To the order $O(\theta)$, the commutator reads:
\begin{eqnarray}
[\partial_{\nu}\phi_0(x)e^{\frac{1}{2} \partial \wedge P},: e^{\pm i\beta\phi_0(y)}e^{\frac{1}{2} \partial \wedge P}:]& =& [\partial_{\nu}\phi_0(x), :e^{\pm
i\beta\phi_0(y)}:]\nonumber\\
&+&\frac{1}{2} [\partial_{\nu}\phi_0(x),:e^{i\beta\phi_0(y)}:] (\overleftarrow{\partial^x}+\overleftarrow{\partial^y})\wedge P\nonumber\\
&-&\frac{i}{2} [\partial_{\nu}\phi_0(x),:e^{i\beta\phi_0(y)}:] (\overleftarrow{\partial^x}\wedge\overleftarrow{\partial^y})
\end{eqnarray}
Now we shall compute the commutator involving twisted fermionic field operators upto the order $\theta$ using:
\begin{eqnarray}
\psi(x) &=&\psi_0(x)e^{\frac{1}{2} \overleftarrow{\partial} \wedge P}\nonumber\\
\bar\psi(x)*\gamma_{\mu}\psi(x) & =& \bar\psi_0(x)*\gamma_{\mu}\psi_0(x)e^{\frac{1}{2} \overleftarrow{\partial}l \wedge P}
\end{eqnarray}
Here,
\begin{eqnarray*}
[\bar\psi(x)*\gamma_{\mu}\psi(x),\frac{1}{2}Z\bar\psi(y)*(1\pm\gamma_{5})\psi(y)]& =& [\bar\psi_0(x)\gamma_{\mu}\psi_0(x),\frac{1}{2}Z\bar\psi_0(y)(1\pm\gamma_{5})\psi_0(y)]\\
&+&[\bar\psi_0(x)\gamma_{\mu}\psi_0(x),\frac{1}{2}Z\bar\psi_0(y)(1\pm\gamma_{5})\psi_0(y)](\overleftarrow{\partial^x}+\overleftarrow{\partial^y})\wedge P\\
&-&\frac{i}{2} [\bar\psi_0(x)\gamma_{\mu}\psi_0(x),\frac{1}{2}Z\bar\psi_0(y)(1\pm\gamma_{5})\psi_0(y)] (\overleftarrow{\partial^x}\wedge\overleftarrow{\partial^y})
\end{eqnarray*}
The commutators \cite{Sid} on commutative spacetime read:
$$
[\partial_{\nu}\phi_0(x), :e^{\pm
i\beta\phi_0(y)}:]
 = \pm\beta g_{\nu 0}\delta(\vec x-\vec y):e^{\pm
i\beta\phi_0}:
$$
$$
[\bar\psi_0 (x)\gamma_{\mu}\psi_0 (x),\frac{1}{2}Z\bar\psi_0 (y)(1\pm\gamma_{5})\psi_0 (y)]= \pm2(1+g/\pi)^{-1}\epsilon^{\mu\nu}\beta g_{\nu 0}\delta(\vec
x-\vec y)\frac{1}{2}Z\bar\psi_0(1\pm\gamma_{5})\psi_0
$$
We notice that the equality 
$$\bar\psi(x)*\gamma^{\mu}\psi(x) = - \frac{g}{2\pi}\epsilon^{\mu\nu}\partial_{\nu}\phi(x)$$ will hold to the order $\theta$ provided
the fermionic and the bosonic momentum field operators are the same since
$\frac{1}{2}Z \bar\psi(x)*(1\pm \gamma^5)\psi(x)=
\frac{1}{2}:e_*^{\pm i\beta\phi(x)}:$.



\end{document}